\documentclass[twocolumn,showpacs,amsmath,amssymb,floatfix,superscriptaddress]{revtex4}
\usepackage{graphicx}
\usepackage{dcolumn}
\usepackage{bm}
\usepackage{hyperref} 
\usepackage{latexsym}

\def\l{\ell}
\def\lab{\ell_{\rm AB}}

\def\a{\alpha}
\def\b{\beta}
\def\g{\gamma}

\def\lsim{\mathrel{\raise.34ex\hbox{$<$\kern-.8em\lower1ex\hbox{$\sim$}}}}

\begin{document}
 
\title{Large-Scale Simulations of Diffusion-Limited $n$-Species Annihilation}
\author{Dexin Zhong} 
\author{Roan Dawkins} 
\author{Daniel ben-Avraham}\email{benavraham@clarkson.edu}
\address{Department of Physics, Clarkson University, Potsdam, New~York
13699-5820, USA} 
\date{\today}

\begin{abstract} 
We present results from computer simulations for diffusion-limited $n$-species
annihilation, $A_i+A_j\to0$ $(i,j=1,2,\dots,n;\>i\neq j)$, on the line, for
lattices of up to
$2^{28}$ sites, and where the process proceeds to completion (no further
reactions possible), involving up to $10^{15}$ time steps.  These enormous
simulations are made possible by the renormalized reaction-cell method (RRC). 
Our results suggest that the concentration decay exponent for
$n$ species is $\a(n)=(n-1)/2n$ instead of
$(2n-3)/(4n-4)$, as previously believed, and are in agreement with recent
theoretical arguments~\cite{tauber}.  We also propose a scaling relation for
$\Delta$, the correction-to-scaling exponent for the concentration decay;
$c(t)\sim t^{-\a}(A+Bt^{-\Delta})$. 
\end{abstract}

\maketitle

\section{introduction}
Diffusion-limited reactions have attracted much interest in recent
years~\cite{reactions,kang85}.  The kinetics of such systems is dominated
by local fluctuations in the concentration of the reactants, thus posing a
formidable problem which has not yet been solved: there
exists no comprehensive theoretical approach for the analysis of
diffusion-limited processes.  

Few select models are amenable to exact analysis.  These include one-species
annihilation, $A+A\to0$, and two-species annihilation $A+B\to0$
(see~\cite{dbook} and references therein).  In one dimension, the particle
density for one-species annihilation decays as
$c(t)\sim t^{-1/2}$, while for two-species annihilation (with equal initial
concentrations and same diffusion constants for the two species) $c(t)\sim
t^{-1/4}$.  In either case the result is strikingly different from the
mean-field kinetics of the corresponding reaction-limited process, $c(t)\sim
1/t$.  To bridge the gap between  these disparate behaviors, ben-Avraham and
Redner~\cite{dba} proposed the n-{\it species annihilation} model, where
particles belonging to the
$n$ species $A_1,A_2,\dots,A_n$ diffuse on the line and react immediately uppon
encounter, according to the scheme:
\begin{equation}
\label{react}
A_i+A_j\to 0,\qquad i,j=1,2,\dots,n,\quad i\neq j\;.
\end{equation} 
For $n=2$ we recover two-species annihilation, while in the limit $n\to\infty$
encounters between like-particles are improbable and the model is
equivalent to one-species annihilation.  For intermediate values of $n$, one
expects $c(t)\sim t^{-\a(n)}$.  

In~\cite{dba} it was proposed, following a heuristic scaling argument and
treating fluctuations via the Van Kampen
$\Omega$-expansion~\cite{vankampen}, that
$\a(n)=(2n-3)/(4n-4)$.  This was supported by numerical simulations of lattices
of typically $10^6$ sites, and up to $10^6$ time steps. (In one time step, all
of the particles in the system move one lattice spacing each, on average.) 
Recently, we have
conducted extensive numerical simulations~\cite{dexin}, following the method
of Renormalized Reaction-Cells (RRC)~\cite{benavraham87,dawkins00,shafrir}. 
The systems involved are up to
$2^{28}\approx2.7\times10^8$ sites long, and the processes were simulated to
completion (until no further reactions are possible), for up to $10^{15}$ time
steps. The new data leads us to the conjecture that
$\a(n)=(n-1)/2n$~\cite{dexin}.  We also find a correction to the main decay
mode, of the form $c(t)\sim t^{-\a(n)}(A+Bt^{-\Delta(n)})$, $\Delta(n)=1/2n$.
The same results were found, independently (and unbeknownst to us), by
Deloubri\`ere et al.,~\cite{tauber}.  In their theoretical derivation, they
consider a simplified version of $n$-species annihilation, where domains of
alternating species loose particles to reactions at one and the same rate, in
a synchronous fashion.  The approximation is more than reasonable, yet it
does not rigorously apply to the original model, and analysis of corrections
is certainly beyond its scope.  Moreover, the simulations in~\cite{tauber}
are comparable in size to those in~\cite{dba}.  In what follows, we report
the results of our large-scale simulations, which strongly support the
conclusions of~\cite{tauber}.  We also propose a scaling relation for the
correction exponent $\Delta$ for $n$-species annihilation, and possibly for
other reaction models where particles segregate into distinct domains.

\section{scaling}
As is well known, local fluctuations in the concentrations of the different
species drive the kinetics of $n$-species
annihilation~\cite{kang85,dbook,dba}.  An initially random homogeneous
distribution of the particles evolves into a continuously growing mosaic of
domains of alternating surviving species.  Two lengthscales characterize the
emerging distribution and dominate the system: the inter-domain distance ---
the distance between the last particle in a domain and the first particle in
the domain next to it ---
$\lab(t)$, and the domain length, $\l(t)$~\cite{redner}.  These quantities
grow with time as
\begin{equation}
\label{expsbg}
\l(t)\sim t^{\b},\qquad\lab(t)\sim t^{\g}.
\end{equation}
Once domains form, reactions might take place only at the domain boundaries,
and particles have to diffuse across the domain gap $\lab$ to react with
other species.  This takes a typical time of $\Delta t\sim \lab^2/D$, where
$D$ is the diffusion constant.  The change in particle concentration during
time $\Delta t$ equals the total number of domain boundaries divided by the
lattice size $L$; $\Delta c\sim -(L/\l)/L=-1/\l$. Thus,
\begin{equation}
\label{dc/dt}
\frac{\Delta c}{\Delta t}\sim-\frac{D}{\l\lab^2}.
\end{equation}  
On substituting the
relations~(\ref{expsbg}) and
$c(t)\sim t^{-\a}$, we derive the scaling rule
\begin{equation}
\label{scaling}
2\g+\b-\a=1.
\end{equation}
Due to the underlying transport mechanism, we expect that
domains grow diffusively, as $\l\sim t^{1/2}$, so $\b=1/2$, and in effect
there is only one independent exponent: $2\a-\g=1/2$.  The general scaling
form holds also for two-species annihilation in the presence of drift (and
with ard-core repulsion between like species), where $\a=1/3$, $\beta=7/12$,
and $\g=3/8$~\cite{shafrir}.

    \section{simulation results}
The $n$-species annihilation process is simulated as follows.  The sites of
a one-dimensional lattice are either empty or occupied by a  
particle (of one of the $n$ species).  Periodic boundary conditions are
imposed, so the lattice is effectively a ring.  At each Monte Carlo step a
particle is chosen randomly and is moved to the nearest site to its right or
left, with equal probabilities.  If the target site is occupied by a particle
of a different species, then both particles are removed from the system,
mimicking the reaction~(\ref{react}).  If the target site is
occupied by a particle of the same species, then the move is disallowed and
it does not take place.  Regardless of the outcome, time is incremented by
$1/N(t)$, where $N(t)$ is the total number of extant particles.  

As the simulation proceeds, the particle concentration declines and the
typical distance between particles increases.  The time spent on simulating
the diffusive motion of the particles until they interact grows even
faster, as the square of the distance between them.  Because of that, computer
simulations are limited to relatively short times.  This problem is overcome
by the RRC method~\cite{benavraham87,dawkins00,shafrir}.

In the RRC method the particles occupy cells, rather
than sites.  Each time the concentration halves, the cells are
renormalized: every two cells are merged into one, and time is renormalized
accordingly. The typical time required to diffuse out of a renormalized cell
twice as large as that of the previous generation is four times longer. 
Thus, physical time is simulated faster with each renormalization step and
the process can be simulated to completion.  Other details for the
implementation of the RRC method are discussed in~\cite{shafrir}. 

Simulations were performed on DEC
Alpha processors running Linux.  Since each lattice site requires 6 bytes
(for species, number of particles, and a pointer to a list of populated sites
that is used for fast selection at each Monte Carlo step), with 2 Gigabytes
($2^{31}$ bytes) memory we were able to simulate lattices of up to $2^{28}$
sites.  The compiler was given special $\#$pragma pack(1) instructions to
circumvent word alignment (which would allocate $32$ bytes for our $6$-byte
site).

\begin{figure}[ht]
 \vspace*{0.cm}
 \includegraphics*[width=0.35\textwidth]{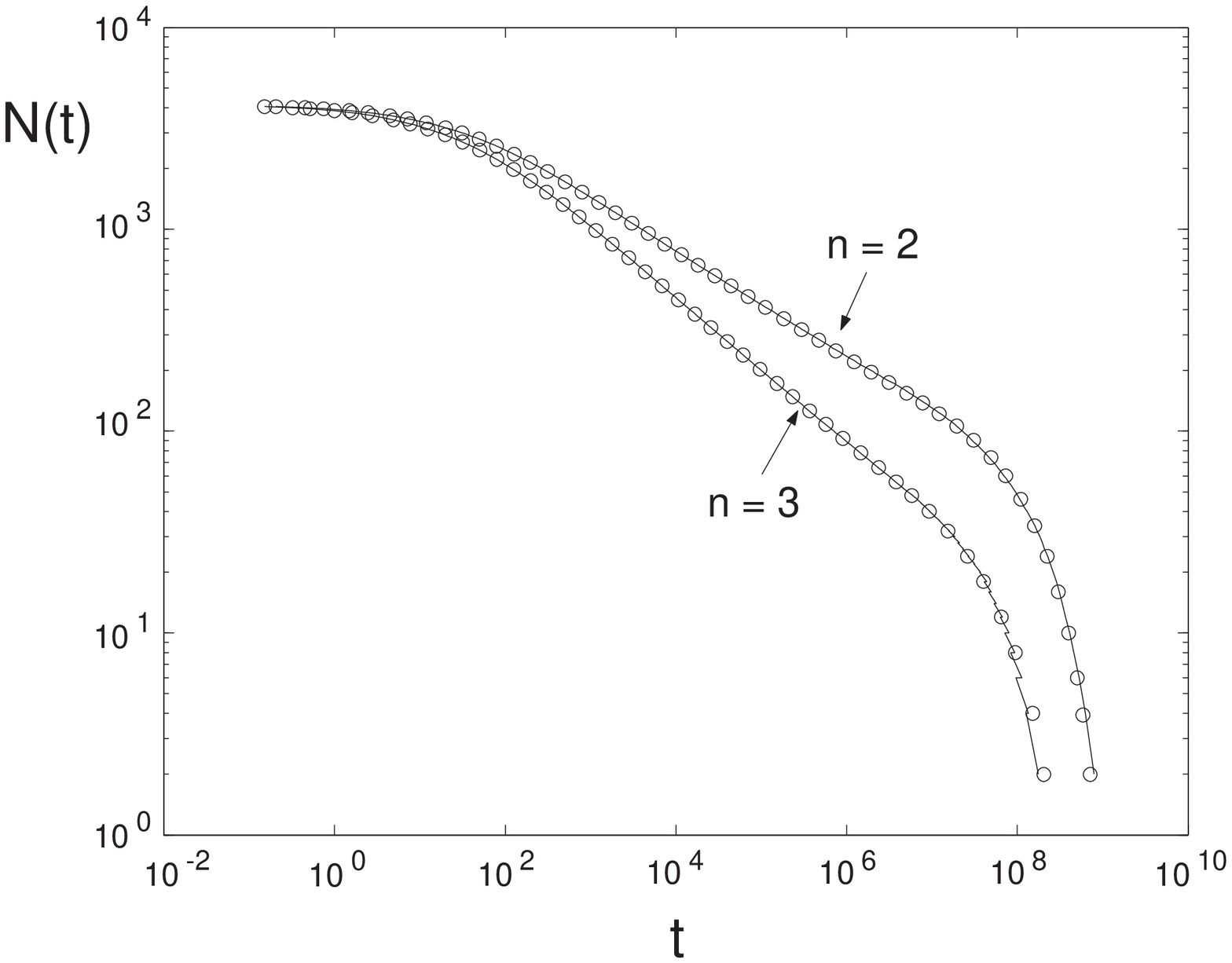}
 \vspace*{0.6cm}
 \includegraphics*[width=0.35\textwidth]{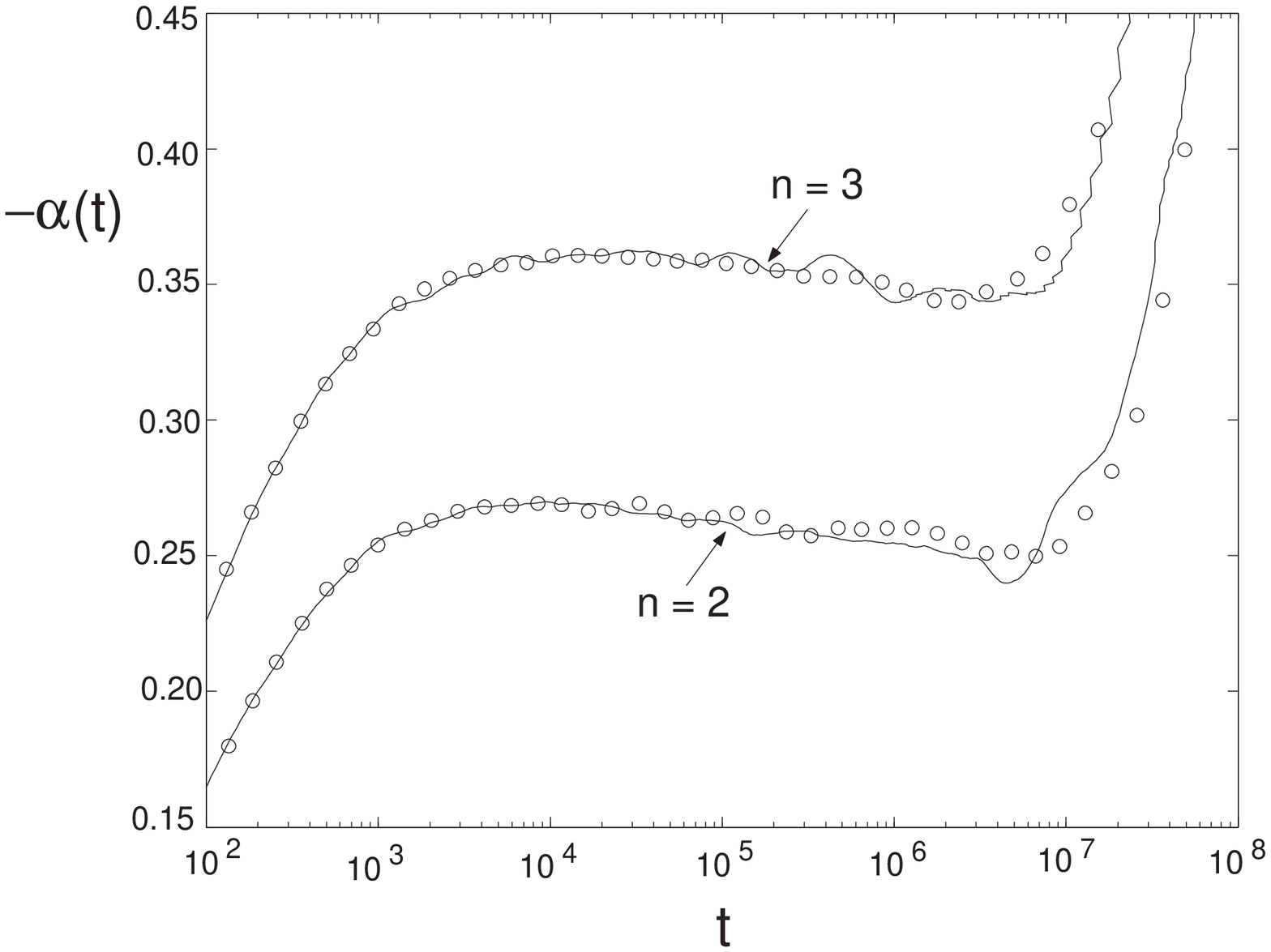}
\caption{Comparison between the traditional (circles) and RRC (solid line)
simulation methods. Plotted is the number of surviving particles $N(t)$
(top) and the local slope $\a(t)=-d\ln N(t)/d\ln t$ (bottom) for
$2^{16}$-site lattices.}
\label{test}
\end{figure}

To test the technique, we have simulated the cases of $n=2$ and $n=3$ on
lattices of $2^{16}=65,536$ sites, in both the RRC and the traditional
simulation method.  These lattices are small enough to enable the simulation
of the process by the traditional method to completion.  On the other hand,
the system is large enough to let us examine the effect of the
renormalizations: with $2^{16}$ sites and $c(0)=1/16$ the RRC method
requires $12$ renormalizations.  In Fig.~\ref{test}a we compare the particle
concentration as obtained by the two methods.  In Fig.~\ref{test}b we plot
the local slope of the curves, the exponent $\a(t)$.  The renormalizations
are discernable only in this second, more stringent test, but the overall
agreement is excellent.  Similar results were obtained for the domain size
and the distance between domains.

\begin{figure}[ht]
 \vspace*{0.cm}
 \includegraphics*[width=0.35\textwidth]{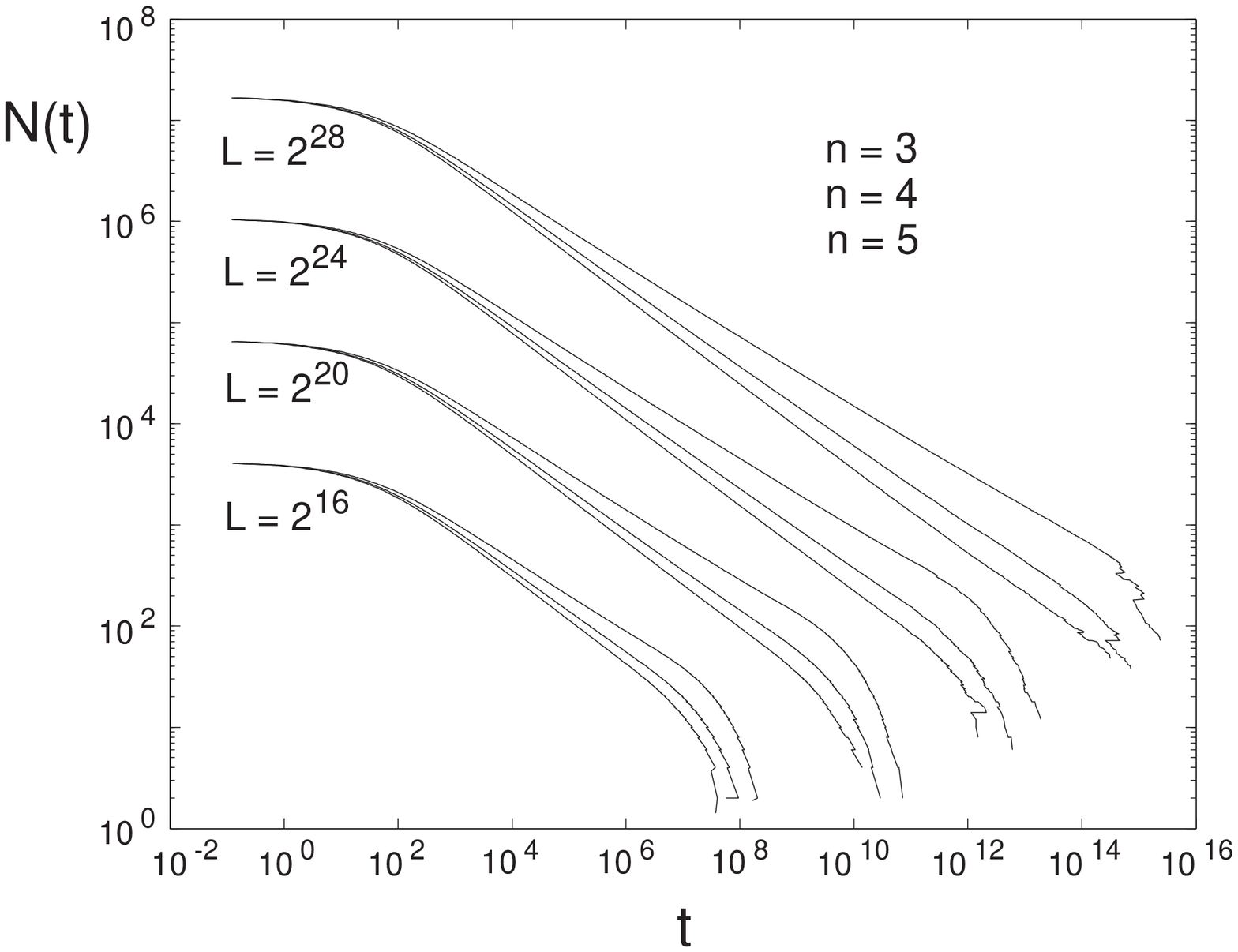}
\caption{Concentration decay for $n=3$, $4$, and 5-species annihilation. 
Plotted is the number of surviving particles, N(t), for system sizes
$L=2^{16}$, $2^{20}$, $2^{24}$, and $2^{28}$ (bottom to top).}
\label{N}
\end{figure}

\begin{figure}[ht]
 \vspace*{0.cm}
 \includegraphics*[width=0.35\textwidth]{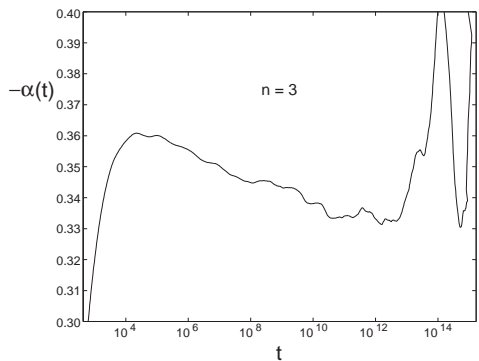}
\caption{Local decay exponent for 3-species annihilation.}
\label{alpha}
\end{figure}

\begin{figure}[ht]
 \vspace*{0.cm}
 \includegraphics*[width=0.35\textwidth]{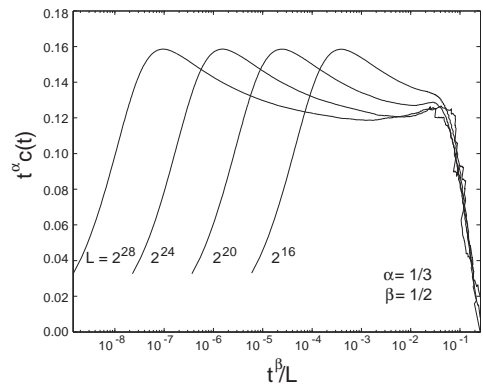}
\caption{Scaling of concentration, $c(t)=t^{-\a}\rho(t^{\b}/L)$, for
3-species annihilation.  The best data collapse (at late times) is obtained
for
$\a=1/3$ and $\b=1/2$.}
\label{collapse}
\end{figure}

Having gained some confidence in the RRC method, we proceed to 
larger simulations.  In Fig.~\ref{N} we show the surviving number of
particles,
$N(t)$, at time $t$, for $n=3$, $4$, and $5$, and several lattice sizes. 
In Fig.~\ref{alpha}, we plot the local decay exponent $\a(t)$ for our
largest simulations of $n=3$.  The maximum of the curve at
$t\approx10^4$ agrees with the earlier prediction that $\a=3/8$~\cite{dba}. 
(Indeed, simulations in~\cite{dba} yielded a somewhat smaller value than the
theoretical
$3/8$, in perfect agreement with current results.)  However, $\a(t)$ is seen
to diminish with time, suggesting a long-time asymptotic limit of
$\a\approx1/3$.  This limiting value is confirmed in the data collapse
(especially at long times) of Fig.~\ref{collapse}, where we plot $t^{\a}c(t)$
vs.~$t^{\b}/L$ for various system sizes, and $\a=1/3$, $\b=1/2$. 
Independent measurements show that $\b=1/2$, as assumed, to within $2\%$,
and the data collapse of Fig.~\ref{collapse} deteriorates with other choices
for the values of $\a$ and $\b$.

We have analyzed in this fashion $n=3$, $4$, and 5-species annihilation, and
measured the exponents $\a$, $\b$, and $\g$.  Our results are summarized in
Table~\ref{table}.  In all cases, the scaling relation~(\ref{scaling}) seems
to hold, and $\b=1/2$ to within numerical errors.  Looking for a simple
expression of these results, that would have the appropriate limits for the
known cases of $n=2$ (two-species annihilation) and $n\to\infty$
(one-species annihilation), we were led to the conjecture~\cite{dexin}
\begin{equation}
\label{conj}
\a=\frac{n-1}{2n},\quad\b=\frac{1}{2},\quad\g=\frac{2n-1}{4n},
\end{equation}
a result derived independently by Deloubri\`ere et al.~\cite{tauber}

\begin{table}[h]
  \centering
  \begin{tabular}{cccccc}
    \hline\hline
     $n$    &  $\alpha$ & $\frac{n-1}{2n}$ &
$\beta$ &  $\gamma$ & $\frac{2n-1}{4n}$ \\ \hline 
3 & \,0.33(1)\, & \,0.333\, & \,0.50(1)\, & \,0.42(2)\, & \,0.417\,\\ 
4 & 0.39(2)     & 0.375     & 0.50(1)     & 0.44(1)     & 0.434\\ 
5 & 0.42(2)     & 0.400     & 0.50(1)     & 0.47(2)     & 0.450\\ 
\hline \hline
  \end{tabular}
  \caption{Exponents $\alpha$, $\beta$ and $\gamma$}
\label{table}
\end{table}

\begin{figure}[ht]
 \vspace*{0.cm}
 \includegraphics*[width=0.35\textwidth]{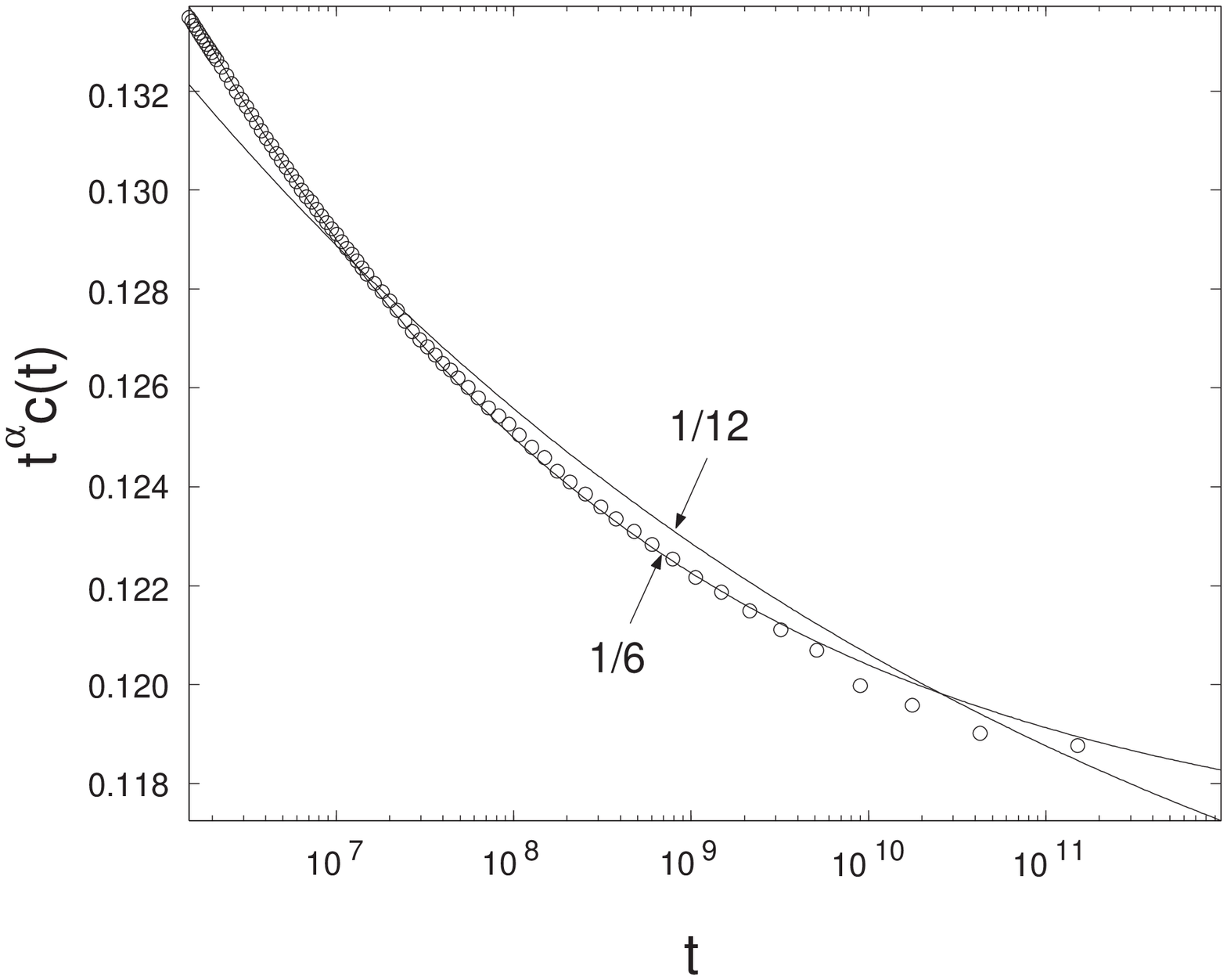}
\caption{Corrections to scaling. Simulation results (circles) are best fit by
Eq.~(\ref{corrections}), with $\Delta=1/6$ (solid line). }
\label{fit}
\end{figure}

Finally, let us address the issue of corrections to scaling of the
concentration decay.  We look for corrections of the form
\begin{equation}
\label{corrections}
c(t)\sim t^{-a}(A+Bt^{-\Delta})\;,
\end{equation}
where $A$ and $B$ are constants. 
Our strategy consists of performing a least squares linear fit of
$A+Bt^{-\Delta}$ to $t^{\a}c(t)$, for different powers $\Delta$, and
searching for the value of $\Delta$ which minimizes the error.  The
scaling form~(\ref{corrections}) is expected to work only after the
asymptotic regime sets in, and before finite-size effects begin, and the
sticky part of our procedure is deciding which times demarcate this region. 
Experimenting with different choices gives us a feel for the errors involved.
In Fig.~\ref{fit} we show best fits
for the region $t=10^6$ -- $10^{12}$, for $n=3$ on a $L=2^{28}$ lattice,
where our data is most reliable.  The results are most compatible with
$\Delta=1/6$ (for $n=3$).  Similar tests for other values of $n$ lead us to
the conjecture that $\Delta(n)=1/2n$.  

The correction exponent can be understood by a simple-minded
argument.  In deriving~(\ref{dc/dt}) we have assumed that the typical
distance between reacting particles, at the edges of adjacent domains, is
$\lab$.  While this is correct, we note that, had the 
distribution of particles been homogeneous, the distance between reacting
pairs would be typically
$\l_{AA}\sim L/c\sim t^{\a}$, quite different from the assumed
$\lab\sim t^{\g}$.  Using $\Delta t\sim \l_{AA}^2/D$ in~(\ref{dc/dt}),
instead of $\lab^2/D$, yields a faster decay; $c\sim t^{-(1-\b)}$. 
Diffusion provides a natural drive toward a homogeneous distribution, and so
it is conceivable that this faster mode of decay is manifested as a
correction to the main behavior, $c\sim t^{-a}$.  It follows
from~(\ref{corrections}) that the correction exponent is
\begin{equation}
\Delta=1-\b-\a=\frac{1}{2n},
\end{equation}
where the last equality applies to $n$-species annihilation, provided that
the conjecture~(\ref{conj}) holds.  The more general relation works well for
two-species annihilation with drift, where $\a=1/3$, $\b=7/12$, and
$\Delta=1/12$~\cite{shafrir}.

\section{Summary and discussion}
We have presented large scale simulation results for diffusion-limited
$n$-species annihilation, in one dimension, using the RRC
method.  Our simulations contradict previous work~\cite{dba} and are in favor
of new theoretical arguments advanced by Deloubri\`ere et al.~\cite{tauber}.
We have also provided a new scaling relation for the correction-to-scaling
exponent $\Delta$, valid for diffusion-limited reactions in one
dimension, where the particles segregate into alternating domains.  The
corrections to the main decay mode are large, and explain the failure
of~\cite{dba} to obtain the correct asymptotic behavior with the size of
simulations employed at that time.  An important conclusion to be drawn is
that predicting asymptotic behavior from the typical size of simulations
used commonly in the field is dangerous.  More advanced techniques and larger
simulations seem to be imperative.

\acknowledgments
We thank S.~Redner for numerous illuminating discussions. We
gratefully acknowledge the NSF (PHY-0140094) for partial support of
this work.


\end{document}